# Avoiding Black hole and Cooperative Black hole Attacks in Wireless Ad hoc Networks

Abderrahmane Baadache
Laboratory of Industrial Technology and Information,
University of A. Mira, Targua Ouzemour, 06000,
Bejaia, Algeria.
.

Ali Belmehdi
Laboratory of Industrial Technology and Information,
University of A. Mira, Targua Ouzemour, 06000,
Bejaia, Algeria.
.

*Abstract*— In wireless ad hoc networks, the absence of any control on packets forwarding, make these networks vulnerable by various deny of service attacks (DoS). A node, in wireless ad hoc network, counts always on intermediate nodes to send these packets to a given destination node. An intermediate node, which takes part in packets forwarding, may behave maliciously and drop packets which goes through it, instead of forwarding them to the following node. Such behavior is called black hole attack. In this paper, after having specified the black hole attack, a secure mechanism, which consists in checking the good forwarding of packets by an intermediate node, was proposed. The proposed solution avoids the black hole and the cooperative black hole attacks. Evaluation metrics were considered in simulation to show the effectiveness of the suggested solution.

## I. INTRODUCTION

A wireless ad hoc network is a collection of nodes connected by wireless links. Although it offers the advantage of being easy to deploy, the wireless ad hoc network paradigm, characterized by the absence of any control at the routing operation or data forwarding, introduced truths problems of security making thus less powerful the operation of such network.

Early conceived routing protocols suppose saint the environment in which the network is deployed [1]. However, that it is the contrary in practice, and it proves to be difficult the guarantee of the security within a network, where the medium of communication is open, a central authority of certification misses, what facilitates the interception, the modification or even the manufacture of packets so that then to inject them into the network, in order to disturb the correct operation or to make entirely non operational the network. Guarantee the security is not limited to ensure, individually, the following services: Participants authentification, data integrity and confidentiality, non repudiation, access control to the communication medium and anonymity. But also, how to put individual solutions together to produce a solution whose range will hold in account the beforehand listed considerations and requirements. The literature contains security solutions which are protocols whose originating objective was security, and other protocols conceived at the beginning without security like primary objective, and they were secured in improved versions because attacks of which they were victims. Obviously, security remains always an arising problem and the remedy is far from being found.

Various attacks against wireless ad hoc networks can be conducted [1]. They are qualified passive ones, if they are limited to the listening of the network traffic to take note, or active if the traffic is modified by the intruder. Security attacks can be internal when the malicious node belongs to the network, or external if not. Deny of service attacks are easy to carry out, and difficult to detect. Their principle is the violation and the non respect of the network protocol specification and their finality is the disturbance of the correct network operation. The no relaying of the traffic (of control or data) by an intermediate node constitutes a behavioral deviation, whose consequence is the violation of the objective for which the network is deployed. Such malicious behavior is called the black hole attack. In this paper and after having specified how a black hole attack is conducted, a solution consisting in checking the good forwarding of the traffic by an intermediate node, was proposed. The solution is based on the well-known principle which is the Merkle tree [2], [4].

The rest of the paper is organized as follows: Section 2 summarizes the related work, follow-up by a necessary background in section 3, then the black hole attack specification is presented in section 4 and detailed description of the suggested solution is the subject of section 5. Simulation and results are analyzed and discussed in section 6. The paper is achieved by a conclusion and our future work.

## II. RELATED WORK

They did not have satisfied solutions to solve the black hole attack problem, what led researchers to be addressed to this attack to find remedies for it. In the literature, the black hole attack solutions can be solutions which are interested in the black hole attack acting in an individual manner or those which are interested in the cooperative black hole attack or general security mechanisms being interested in others attacks in addition to the black hole attack.

In [4], Hongsong et al. proposes an intrusion detection model to combat the black hole attack in AODV [5] routing protocol. In this model, a security agent, established by a hardware thread in network processor uses parallel multithreading architecture, try to detect two cases of figure of attack. Those exploiting AODV control messages RREQ (Route REQuest) and RREP (Route REPly). The agent monitors the RREQ-RREP messages at real-time and if any detection rule is violated, the black hole attack is detected and the malicious node is isolated and recorded to a black list. This solution requires a special material for its implementation. It is dedicated to AODV protocol and it considers only control messages, however that black hole attack can target data messages. Considering always AODV, authors of [6] try to detecting abnormality occurs during the black hole attack by defining a normal state from dynamic training data that is updated at regular time intervals. To express the state of the network, the following features are used: Number of sent out RREQ messages, Number of received RREP messages and the change of the sequence number value used by AODV to determine the route freshness degree. Through the simulation, this method shows significant effectiveness however a more processing overhead is needed for its implementation and consequently it can suffer from scalability problems. In [7], the authors are always interested in AODV and propose a solution in which the receiving node of RREP message compares the sequence number value with a dynamic updated threshold. If the sequence number value is found to be higher than the threshold value, the node is suspected to be malicious and it adds the node to the black list. Here still, except message RREP is controlled. It is necessary to hold in account also data packets because a black hole node can behave normally in the route establishment phase and maliciously in the data transmission phase. In more, the threshold considered can miss exactitude what brings back to false alarms.

A cooperative black hole attack is when several malicious nodes work together as a group. To identify multiple black hole nodes acting in cooperation, Fu's team, in [8], [9], proposes slightly modified AODV protocol by introducing Data Routing Information (DRI) table, that contains information on routing data packet from/through the node and cross checking process that determines the reliable nodes to discover secure paths from source to destination. In [10], authors propose an enhancement of the basic AODV routing protocol to combat the cooperative black hole attack. They use a structure, which they call fidelity table, wherein every participating node will be assigned a fidelity level that acts as a measure of reliability of that node. In case the level of any node drops to 0, it is considered to be a black hole node and is eliminated. In their approach, they assume that nodes are already authenticated which is a little strong assumption. Agrawal et al. [11] proposes a complete protocol to detect a chain of cooperating malicious nodes in an ad hoc network. The proposed protocol is based on sending equal and small sized blocks of data, and monitoring the traffic flow at the neighborhoods of both source and destination, then gathering results of monitoring by a trusted backbone network, with the little strong assumption that a neighborhood of any node has more trusted than malicious nodes.

Other black hole attack solutions were summarized in [1] and dedicated solutions for securing particular routing protocols against all possible attacks, not only against the black hole, were proposed. SAR [12] and SEAD [13] are examples of such secure routing protocols. The watchdog and pathrater [14] mechanism is also proposed for mitigating misbehavior. The watchdog for identifying misbehaving nodes and the pathrater helps routing protocols avoid these nodes. This mechanism is functional only with nodes equipped by interfaces that support promiscuous mode operation.

### III. BACKGROUND

Our solution is based on the principle of Merkle tree [2], [3] and uses the AODV [5] routing protocol like case of study, thus the knowledge of Merkle tree principle and AODV functioning is necessary.

#### A. Merkle tree

A Merkle tree is a binary tree in which, each leaf carries a given value and the value of an interior node (including the root) is a one-way hash function of the node's children values. Figure 1 illustrates an example of a Merkle tree in which:

- *h* denotes a one-way hash function. For example, the function SHA-1 [15].
- *||* is the concatenation operator.
- Values of leaves 1,2,4, respectively, are: *h(a)*, *h(b)*, *h(c)*.
- The value of the interior node 3 is: *h(h(a)||h(b))* which is the hashing result of the concatenation of values of children 1 and 2. Idem for the node 5 whose value is *h(h(h(a)||h(b))||h(c))* and children are: 3 and 4.

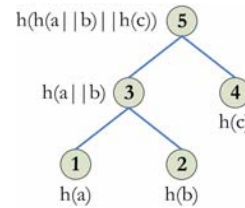

Figure 1. Example of a Merkle tree

#### B. AODV routing *protocol*

AODV (Ad hoc On-Demand Distance Vector) [5] is a reactive routing protocol composed of two modules:

- Route discovery module: To send data to a given destination *D*, the source node *S* consults its routing table. If it finds a valid entry (a route) towards this destination *D*, it uses it immediately, else it launches a route discovery procedure (see Figure 2.), witch consists in broadcasting, by the source node *S*, a route request (RREQ) message (containing amongst other information: destination's address, destination's sequence number) towards neighbors. When RREQ is received by an intermediate node, this last consults its routing table to find a fresh route (the route is fresh if the sequence number of this route is larger than that of RREQ) towards the requested destination in RREQ. If such a route is found, a route reply (RREP) message is sent through the pre-established reverse route (established when RREQ pass through intermediate nodes) towards the source *S*. If the intermediate node does not find a fresh route, it updated its routing table and sends RREQ to these neighbors. This process is reiterated until RREQ reaches the destination node *D*. The destination node *D* sends RREP to *S* by using the pre-established reverse route. It should be noted that the source *S* can receive several RREP, it will choose that whose destination's sequence number is larger, if destination's sequence numbers of several RREP are equal, that of which the smallest hope counter will be selected.

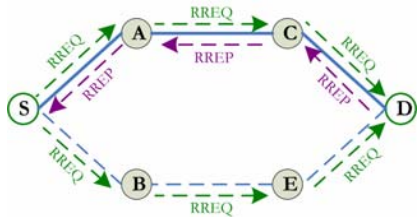

Figure 2.  Route discovery process of AODV

- Route maintenance module: AODV uses Hello messages to maintain the connectivity between nodes. Each node periodically sends a Hello message to these neighbors and awaits Hello messages on behalf of these neighbors. If Hello messages are exchanged in the two directions, a symmetrical link between nodes is always maintained if no link interrupt occurs. The broken link can be repaired locally by the node upstream, else a route error (RERR) message is sent to the source *S* (see Figure 3.). This last can launch again, if necessary, the route discovery procedure. It should be noted that the link interrupt is the consequence of the mobility or the breakdown of nodes.

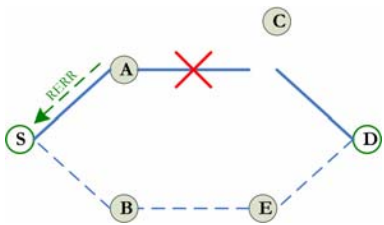

Figure 3.  Route maintenance process of AODV

IV.  BLACK HOLE ATTACK SPECIFICATION

In a black hole attack, the malicious node refuses to forward data packets to the following node in the route connecting a given source and destination. To conduct its attack, the malicious node must initially belong to the data route, then it pass to the action which is the data dropping. According to the specification of the target routing protocol, the manner with which the malicious node fits in the data route differs. Since our case of study is the AODV protocol, we will see how a malicious node can make a success of its attack in AODV.

Two kinds of black hole attack can be distinguished:

- Internal black hole attack: The malicious node is an internal node which does not seek to fit in an active route between a given source and destination, and if the chance would have it, this malicious becomes element of an active data route, it will be able to conduct its attack as the transmission of the data starts. This attack is internal because the malicious node belongs already to the data route. Here, there is no violation of AODV specification and the malicious node has anything to make to carry out with success its attack.

- External black hole attack: The malicious node is an external node which seeks to fit in an active route. For that, it violates the routing protocol specification and executes the process schematized in Figure 4. and summarized in the following points:
  - The malicious node detects the existence of an active route and takes note of the destination address.
  - The malicious node prepares a route replay packet (RREP) in which: the destination address field is set to the spoofed destination address, the sequence number is set to a greatest value and the counter hope is set to a smallest value.
  - The malicious node sends this route reply RREP to the nearest intermediate node belonging to the real active route (not necessarily to the data source node himself).
  - The route reply RREP received by the intermediate node will be relayed through the preestablished inverse route towards the data source node.
  - The source node updates its routing table by the new information received in the route reply.
  - The source uses the new route to sending data.
  - The malicious node starts to drop the data in the route to which it belongs.

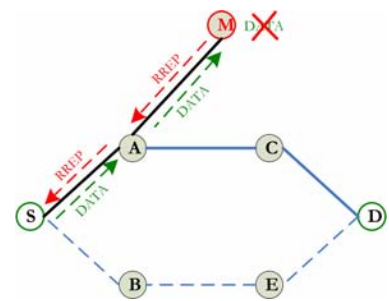

Figure 4.  Black hole attack specification

The malicious node *M* detects the existence of active route (*S,A,E,D*) between the source *S* and destination *D*. The malicious node *M* sends to the node *A* a route reply message (RREP) containing the spoofed destination address, a relatively large sequence number and a small hope count. The node *A* forwards this route reply message to *S*. This last updates its routing table and considers the new route (*S,A,M*). The node *S* uses this route to send data and while arriving at *M*, these data will be quite simply dropped. Nodes source and destination will not be able any more to communicate in the presence of the black hole attack.

## V. OUR SOLUTION

The table I contains the notations used to describe our solution.

TABLE I.  NOTATIONS

| Notation | Significance |
|---|---|
| $id_i$ | Identity of node i. |
| $S_i$ | Secret generated by node i. |
| $h$ | One-way hash function. |
| // | Concatenation operator. |

In Figure 5, we consider a piece of network made up of 3 nodes *A*, *B* and *C*. On this last, a Merkle tree is juxtaposed. We point out that our goal is to check that *B* conveys well, towards *C*, the traffic sent by *A*.

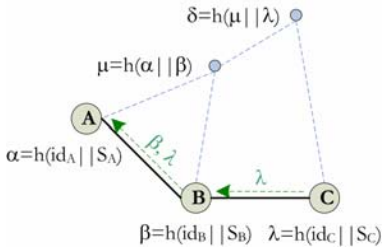

Figure 5.  Basic principle of the solution, single black hole case

Each node *i* holds the value $h(id_i||S_i)$, $i \in \{A,B,C\}$. The transmitter node *A* has, in more, the value δ (value of the root of the Merkle tree). So that *A* checks that *B* forwards well the traffic to *C*, the node *C* sent the value λ (value held by *C*) to *B* and *B*, in turn, sent to *A*, β (value held by *B*) and λ. When both β and λ are received by *A*, the node *A* recalculates δ from α (value owned by A), β and λ, then compares the result with the value of δ already held, if equality, the node *B* routed out the traffic to *C*, otherwise, *B* is a black hole node. Obviously, to leading a black hole attack, the node *B* must generate the value λ which is impossible, because it does not know the *C*'s secret $S_c$.

Nodes *B* and *C* can cooperate to conduct black hole attack, this is easy if *C* communicates to *B* its secret $S_c$. To prevent such a scenario, the idea of the solution can be generalized as it is shown in Figure 6.

When β, λ and ω are received by *A*, the node *A* recalculates ψ from α, β, λ and ω, then compares the result with the value ψ of already held, if equality, the route (*A,B,C,D*) is secured, otherwise, the route contains a black hole nodes. *B* and *C* cannot conduct a cooperative black hole attack because they do not know the secret $S_D$ held by *D*.

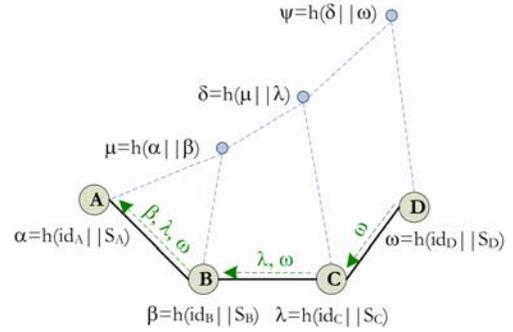

Figure 6.  Basic principle of the solution, cooperative black hole case

It is important to note that the secret $S_i$ is generated by node *i* independently of other nodes and this to prevent another node to replay the role of the node *i*.

Once detected, the black hole node can be put in a black list to avoid it in all future communication and, if necessary, another discovery route operation can be started again. For the implementation of our solution, an initialization process is necessary. It enables to nodes to generate each one its secret and communicate the root value of the Merkle tree to the transmitter node (communicate the value δ to the node *A* in Figure 5).

The gray hole attack is a variant of the black hole attack in which a malicious node, selectively, destroys packets of the traffic that passes through it. Our solution can easily be adapted to thwart such attacks. It is remarkable that this solution can be integrated in any routing protocol, at the time when its operation is not related to any specificity of a particular routing protocol. Obviously, the specification of the black hole attack is different from a routing protocol to another.

## VI. SIMULATION AND ANALYSIS

In the network of the Figure 7, we use a source data node, a destination data node and one or more black hole nodes chosen randomly among nodes of the network. OPNET Modeler version 11.5 [16] is used as a simulator. The table II contains the simulation parameters.

TABLE II.  SIMULATION PARAMETRES

| Simulation parameter | Value |
|---|---|
| Nodes number | 10 |
| Network size | 1km*1km |
| Simulation duration (sec) | 600 |
| Packet Inter-Arrival Time (sec) | exponential(1) |
| Packet size (bits) | exponential(1024) |
| Transmit Power (watt) | 0.0001 |
| Routing protocol | AODV |
| Hash function | SHA-1 |

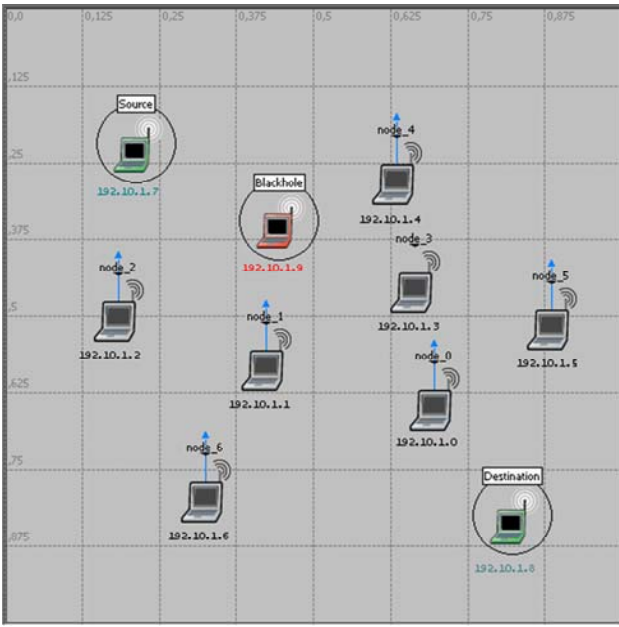

Figure 7. Network used in simulation

In this section, we show the effect of the black hole attack on the network functioning in first place, and how our solution can neutralize such an attack in second. This, in the presence of one or more black hole attacks (cooperative black hole). For this, we consider the following metrics:

- Traffic sent by the node source (packet/sec): indicate the number of packets/second sent by the source node.
- Traffic received by the destination node (packet/sec): indicate the number of packets/second received by the destination node.

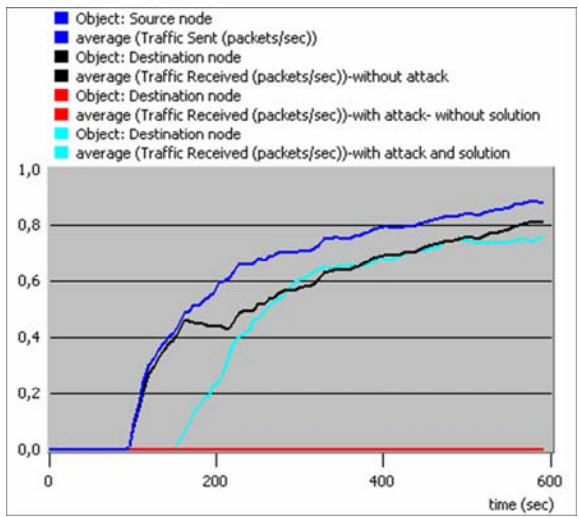

Figure 8. Traffic Sent & Traffic Received - Single Black hole

Figure. 8 shows the traffic sent by the source node, the traffic received by the destination node without attack, the traffic received by the destination node in the presence of a single black hole node and the traffic received by the destination node when the solution is used. The traffic received by the destination node is null under the effect of the attack without using the solution, this is justified by the destruction of packets per black hole node. Under the effect of the solution, le black hole node is eliminated and the destination node receives normally the traffic.

Figure 9. shows the simulation results in the presence of two malicious nodes cooperating together to conduct a black hole attack (cooperative black hole).

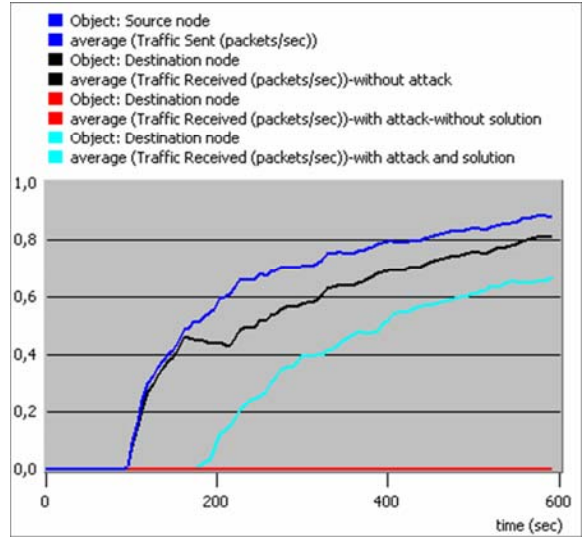

Figure 9. Traffic Sent & Traffic Received - Cooperative Black hole

Our solution detects well the cooperative black hole attack but with a little time lag because the detection of the cooperative black hole attack requires more calculation than in the case of a simple black hole, then an alternative route is found to convey the traffic to the destination node.

To evaluate the performance of our solution, we consider the following evaluation metrics:

- End to end delay (sec): indicate the delay, in second, for sending a bit of the source node to the destination node.
- Network load (bits/sec): indicate the traffic quantity, in bits/sec, in the entire network.

Figure 10. and Figure 11. show, respectively, the end to end delay and the network load, in cases : without black hole attack, with black hole attack and solution, and cooperative black hole attack and solution.

A very small time lag due to tests and calculations carried out by nodes, after which, the delay is stabilized and become almost identical in all scenarios, which shows that our solution does not influence the degradation of the network performance.

A slight increase in the network load because of messages exchanged between nodes to communicate various hash values. More messages in the case of cooperative black hole, comparatively with the case of single black hole, which justified graphs pace in Figure 11.

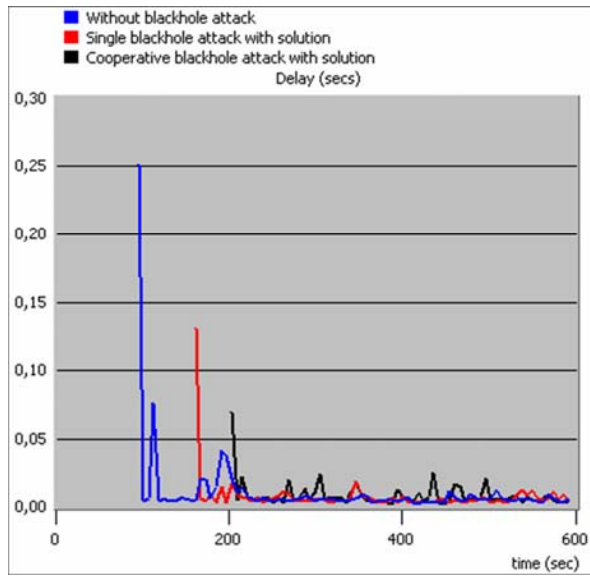

Figure 10. End to end delay

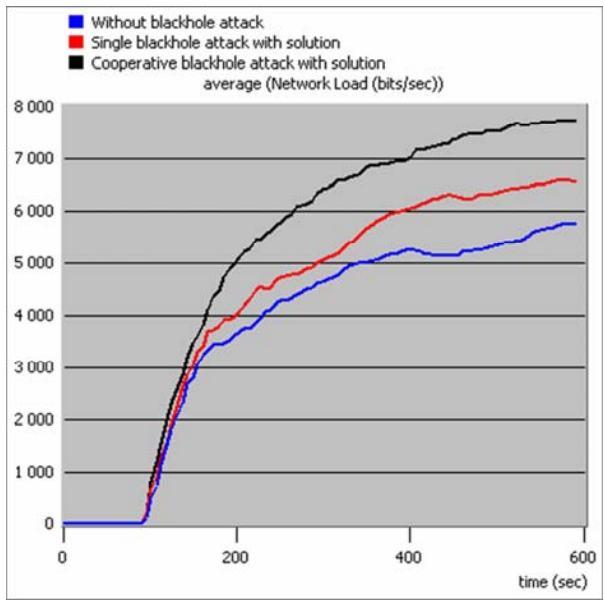

Figure 11. Network load

In the issue of these results, the network performance remains acceptable, if it is only a negligible variation in the end to end delay and a slight increase in the network load.

## VII. CONCLUSION AND FUTUR WORK

Ad hoc network security is a serious problem by which researchers were concerned. Several security solutions are suggested but perfect security is far from beings obvious. We focused in this paper to the black hole attack, which refuses to convey the traffic and drop it. After the black hole attack specification in an example of routing protocol (AODV), we proposed a solution, based on the principle of Merkle tree, for avoiding the black hole and the cooperative black hole attacks.

Moreover, Gray hole, a variant of the black hole, can easily be detectable by our solution. Simulation results show well its effectiveness in the detection of such attacks and the slight influence of this solution on the network performance. It was announced that our solution is general and it is not dedicated to any routing protocol.

We think that the network density, nodes mobility and the number of black hole nodes are determining factors in our solution performance, in term of end to end delay and network load. In a future work, we will study the influence of these factors and will find adequate mechanisms which will make the solution more powerful.

ACKNOWLEDGMENT

Colleagues, H. Moumen and S. Mustapha are thanked for the second reading of this paper and for pertinent criticisms and constructive remarks.

AUTHORS PROFILE

**Abderrahmane Baadache** obtained the *engineer* degree in computer science, option computing systems, from the national institute of computing (INI), Oued Smar, Algiers, Algeria in 2001 and the *magister* degree, option networks and distributed systems from the doctoral school of networking and distributed systems, department of computer science, A. Mira University, Béjaia, Algeria in 2005. Currently, he is preparing his PhD under the direction of the professor A. Belmehdi. His research topic is wireless networking, in particular, security in ad hoc, sensor, mesh and vehicular networks.

**Ali Belmehdi** is a professor at the University of A. Mira, Bejaia, Algeria. Besides his teaching activities, he is the header of the Laboratory of Industrial Technology and Information. His career is rich in scientific publications and supervising students in PhD or magister degree. His research topic includes the automatic and the computer science, particularly, wireless networking.